\documentclass[12pt]{article}
\usepackage{graphicx}

\newcommand\pubnumber{}
\newcommand\pubdate{\today}

\def\prague{On behalf of the ATLAS and CMS Collaborations\\}
\def\support{\footnote{Institute of Physics of the ASCR, vvi\\
Na Slovance 2, CZ-182 21 Praha 8, Czech Republic}}

\def\Title#1{\begin{center} {\Large #1 } \end{center}}
\def\Author#1{\begin{center}{ \sc #1} \end{center}}
\def\Address#1{\begin{center}{ \bf #1} \end{center}}

\newcommand\pubblock{\rightline{\begin{tabular}{l} \pubnumber\\
         \pubdate  \end{tabular}}}
\newenvironment{Abstract}{\begin{quotation}  }{\end{quotation}}
\newenvironment{Presented}{\begin{quotation} \begin{center} 
             PRESENTED AT\end{center}\bigskip 
      \begin{center}\begin{large}}{\end{large}\end{center} \end{quotation}}

\def\beq{\begin{equation}}
\def\eeq#1{\label{#1}\end{equation}}
\def\eeqn{\end{equation}}

\def\beqa{\begin{eqnarray}}
\def\eeqa#1{\label{#1}\end{eqnarray}}
\def\eeqan{\end{eqnarray}}

\let\bar=\overbar

\def\Dslash{\not{\hbox{\kern-4pt $D$}}}
\def\dslash{\not{\hbox{\kern-2pt $\del$}}}

\def\msb{{\bar{\ssstyle M \kern -1pt S}}}

\begin{document}
\begin{titlepage}
\pubblock

\vfill
\Title{$W/Z$ measurements at LHC}
\vfill
\Author{ Pavel Staroba\support}
\Address{\prague}
\vfill
\begin{Abstract}

Production of $W$ and $Z$ bosons in proton - proton collisions at
Large Hadron Collider has been extensively studied by the 
ATLAS and CMS Collaborations
during the Run 1 period. Data collected at $\sqrt{s}=7$ TeV and $\sqrt{s}=8$ TeV
were analysed. Both collaborations produced nearly 60 publications dealing with
$W/Z$ physics in total. A wide range of topics is covered.

A representative selection of the aforementioned results is presented. Expectations
for the Run 2 period are summarised.

\end{Abstract}
\vfill
\begin{Presented}
Twelfth Conference on the Intersections of Particle and Nuclear Physics\\
Vail, Colorado,  May 19--24, 2015
\end{Presented}
\vfill
\end{titlepage}
\def\thefootnote{\fnsymbol{footnote}}
\setcounter{footnote}{0}

\section{Introduction}

  Investigation of production and properties of the $W$ and $Z$ bosons constitutes a significant part of the research
program at hadron colliders. Wide range of possible measurements allows one to investigate
a broad range of aspects. It is possible to perform basic and precision tests of quantum chromodynamics(QCD), 
to constrain parton distribution functions(PDFs),
to test parton emission calculations and resummation models, to do stronger and more stringent tests of QCD, or to extract
Standard Model parameters. Some measurements of $W$ and $Z$ boson properties are sensitive to physics beyond the Standard Model.

During the Run 1 period at LHC~\cite{LHC}, the ATLAS~\cite{ATLAS01} and CMS~\cite{CMS01} Collaborations produced nearly
60 publications dealing with $W/Z$ physics in total. A wide range of topics is covered: 
production cross section of  inclusive $W$ and $Z$  bosons decaying into $e$, $\mu$, $\tau$ and $b$ channels, 
differential Drell-Yan cross sections, 
differential cross sections of $W$ and $Z$ as a function of boson transverse momentum, rapidity and $\Phi^{*}_{\eta}$ observable,
production of  $W$ and $Z$ in association with light jets,
ratio of the $W$ + jets to $Z$ + jets cross sections,
production of  $W$ and  $Z$ in association with $c$ and $b$ quark, 
forward-backward asymmetry of  Drell-Yan lepton pairs,
lepton charge asymmetry in inclusive $W$ production,
weak mixing angle with the Drell-Yan process,
angular distributions in the decay of $W$ and $Z$,
and electroweak production of dijets in association with a $Z$ boson.

To provide as apposite and as transparent as possible representation
of $W/Z$ physics results achieved by both collaborations during the Run 1 period, main
results of selected publications on the following topics will be briefly depicted in the next:
inclusive production cross section of $W$ and $Z$ decaying leptonically, differential Drell-Yan cross section,
differential cross section of $Z$ as a function of boson transverse momentum, rapidity and $\Phi^{*}_{\eta}$ observable,
production of $Z$ in association with light jets,
ratio of the $W$ + jets to $Z$ + jets cross sections and
production of a $Z$ boson in association with a $b$ quark.

\section{Selected results from Run 1}

  The CMS Collaboration performed measurements of inclusive $W$ and $Z$ boson production cross sections
in the lepton channel~\cite{CMS02} using 18 pb$^{-1}$  of 8 TeV data. This data sample
with low pileup was collected in May 2012. Assuming lepton universality, the results from electron and muon
channels are combined by calculating an average cross section value weighted
by their statistical and systematic uncertainties.

\begin{figure}[!ht]
\begin{center}
 \begin{tabular}{cc}
  \begin{minipage}{.5\hsize}
   \begin{center}
     \includegraphics[width=7cm]{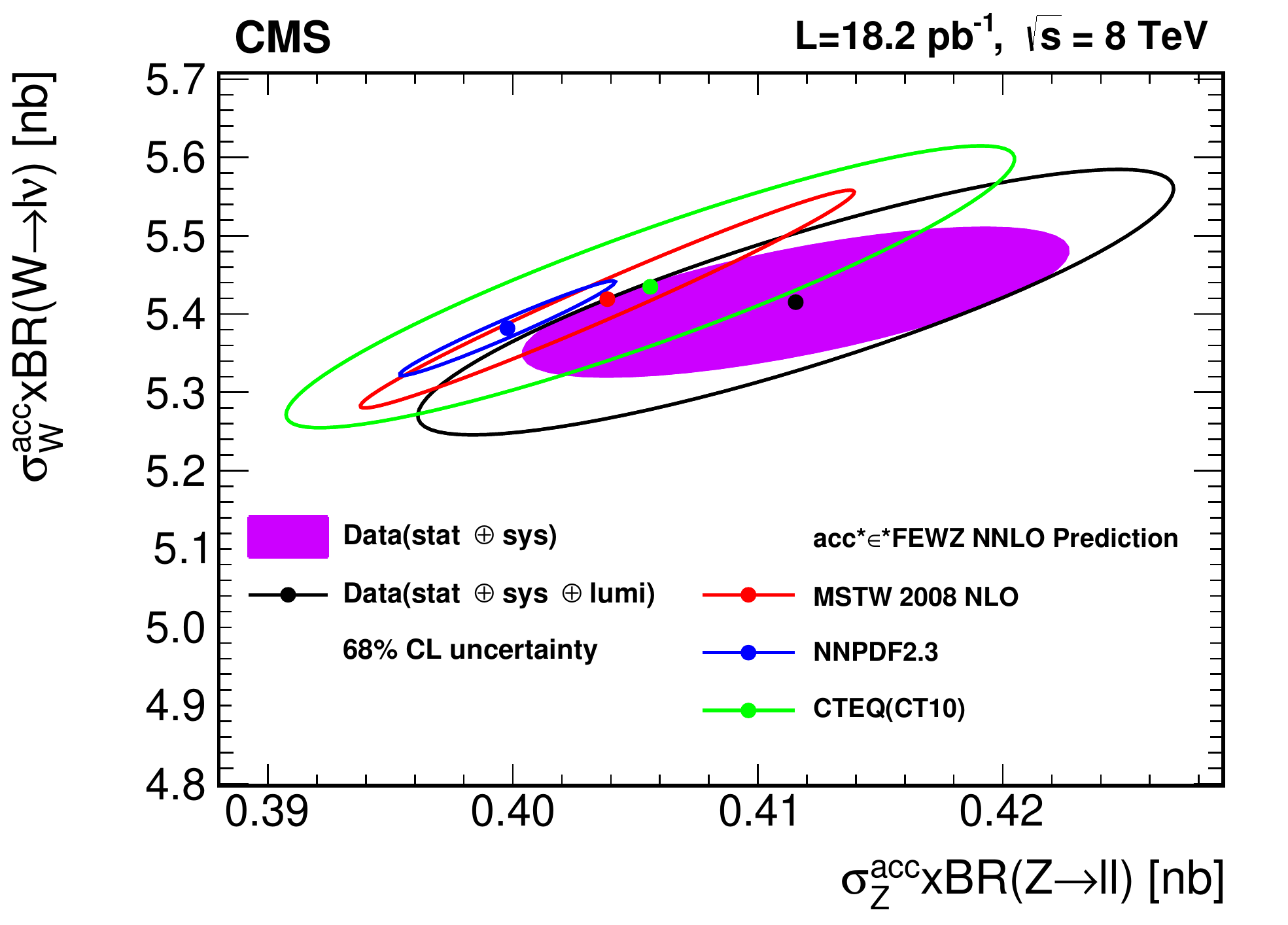}
   \end{center}
  \end{minipage}

  \begin{minipage}{.5\hsize}
   \begin{center}
     \includegraphics[width=7cm]{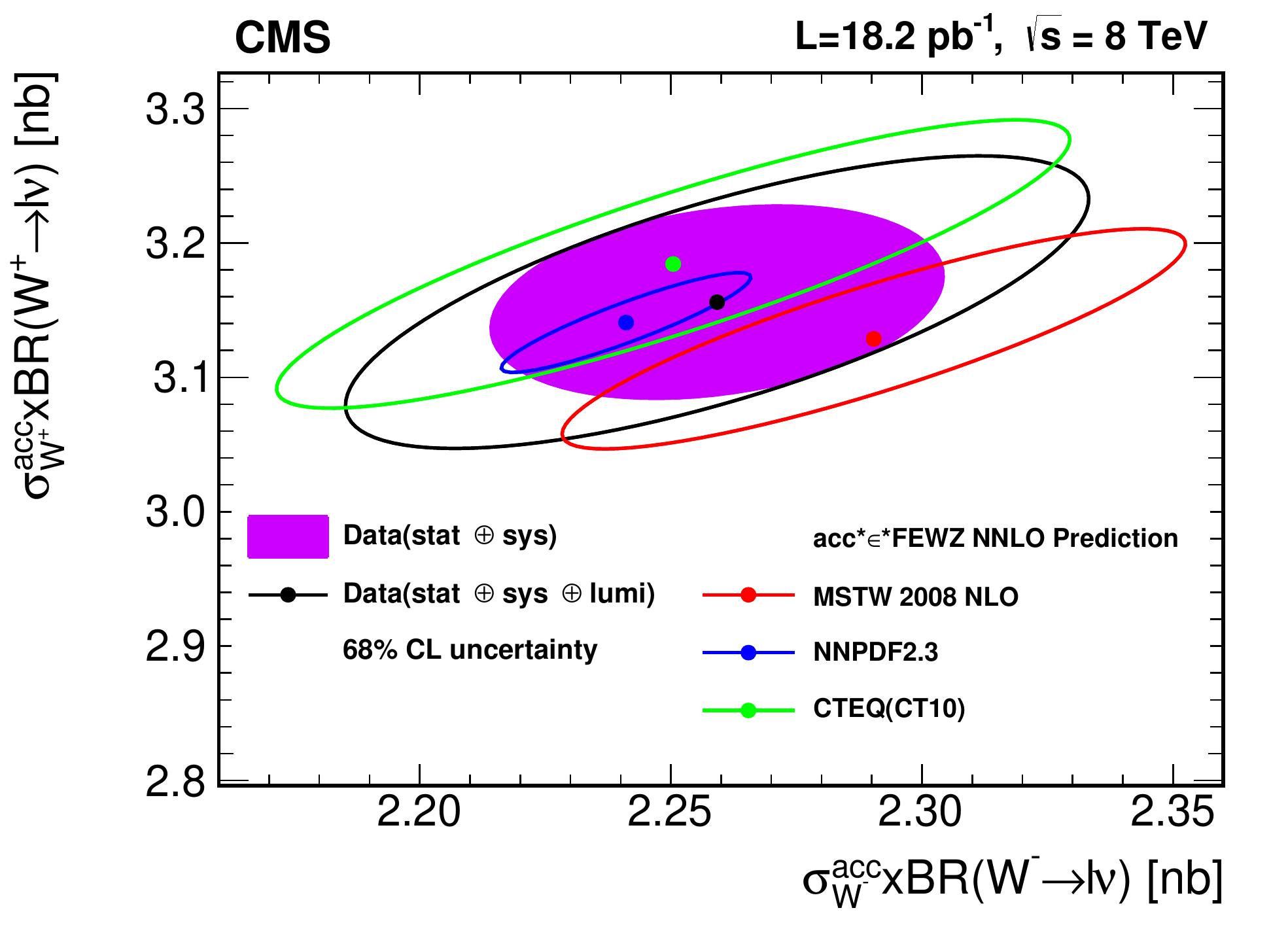}
   \end{center}
  \end{minipage}
 \end{tabular}
\caption{Comparison of measured and predicted fiducial cross sections of $W$ boson
versus $Z$ boson and of $W^{+}$ boson versus $W^{-}$ boson. Taken from Ref.~\cite{CMS02}.
 \label{fig:fig01}}
\end{center}
\end{figure}

  Results are compared with the theoretical prediction calculated at NNLO with
the program Fully Exclusive $W$ and $Z$ Production (FEWZ ~\cite{FEWZ}) using
MSTW2008 set of PDFs~\cite{MSTW2008}. Comparison of measured and predicted fiducial cross sections of $W$
versus $Z$ and of $W^{+}$ versus $W^{-}$ is shown in Figure~\ref{fig:fig01} for several PDF sets. 
The measurement is consistent with theoretical predictions. It has a discriminating power between
PDF sets.

  Drell-Yan differential cross section as a function of dilepton mass was 
measured at 7 TeV by the ATLAS Collaboration and at 7 and 8 TeV by the CMS Collaboration.

The ATLAS Collaboration measured high mass Drell-Yan fiducial differential cross section using 4.9 fb$^{-1}$ 
of 7 TeV data ~\cite{ATLAS02}. Measurement was performed only in the
electron channel. The measured cross section was compared to 
the theoretical prediction calculated using 
FEWZ3.1 with 5 NNLO PDFs. 
Measurement is consistent with the theory prediction, although the data
are systematically above the theory. Measurement does not enable to distinguish
between PDFs.

Low mass Drell-Yan distribution was measured by the ATLAS Collaboration
using 35 pb$^{-1}$ collected in 2010 and 1.6 fb$^{-1}$
collected in 2011 ~\cite{ATLAS03}. The measurement was performed in the electron and muon
channel in the range of dilepton masses between 26 and 66 GeV and these
measurements are combined. The analysis is extended to the range
between 12 and 66 GeV in the muon channel using only data from the
year 2010.
The measured fiducial differential cross section is compared to NLO
prediction (FEWZ), NLO + Leading Logarithm resummed Parton Shower
(POWHEG~\cite{POWHEG}) and NNLO (FEWZ). Only NNLO prediction is able to reproduce
the data. The necessity of using NNLO prediction was confirmed by QCD analysis
of data.

\begin{figure}[!ht]
\begin{center}
 \begin{tabular}{cc}
  \begin{minipage}{.5\hsize}
   \begin{center}
     \includegraphics[width=7cm]{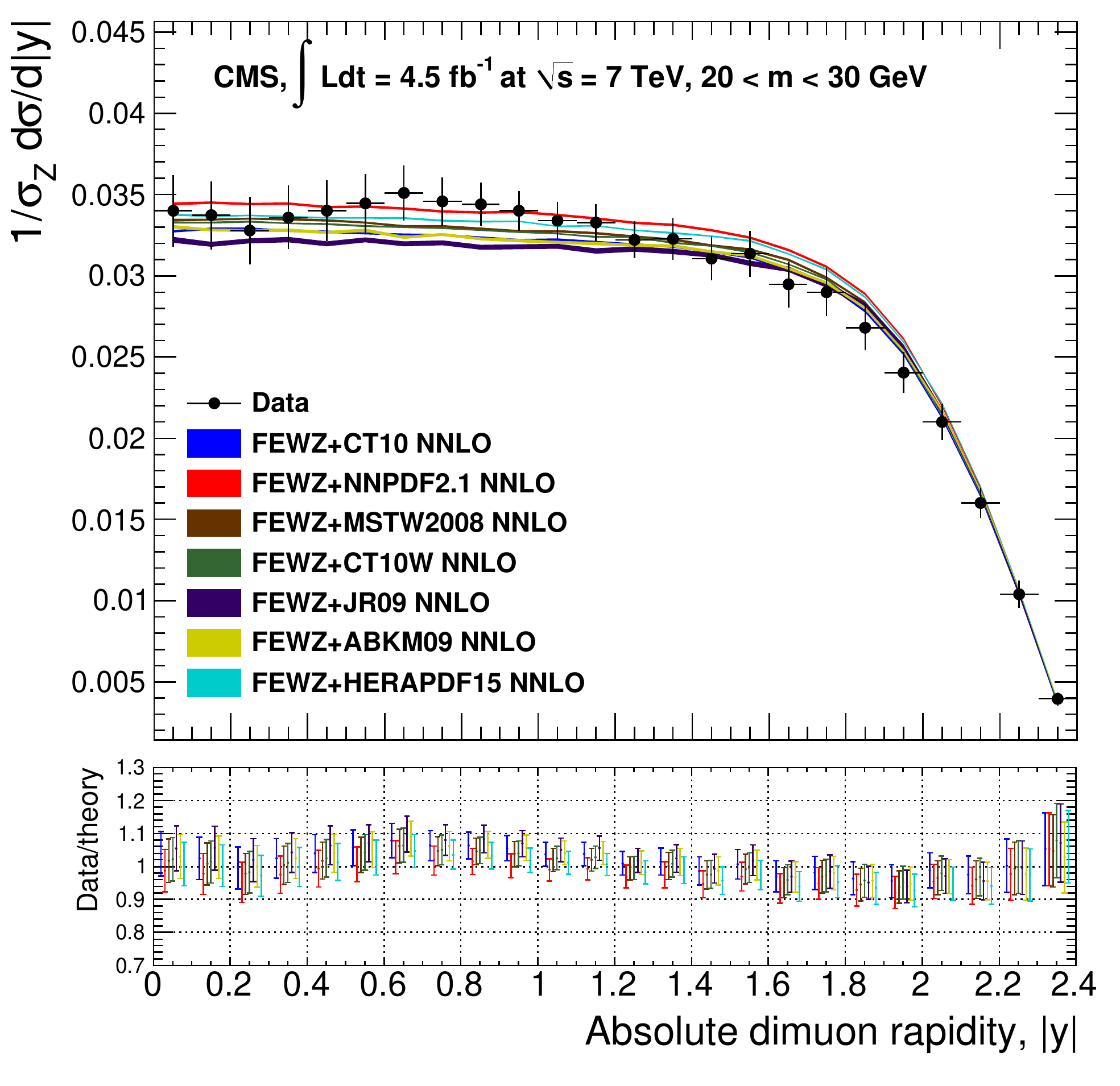}
   \end{center}
  \end{minipage}

  \begin{minipage}{.5\hsize}
   \begin{center}
     \includegraphics[width=7cm]{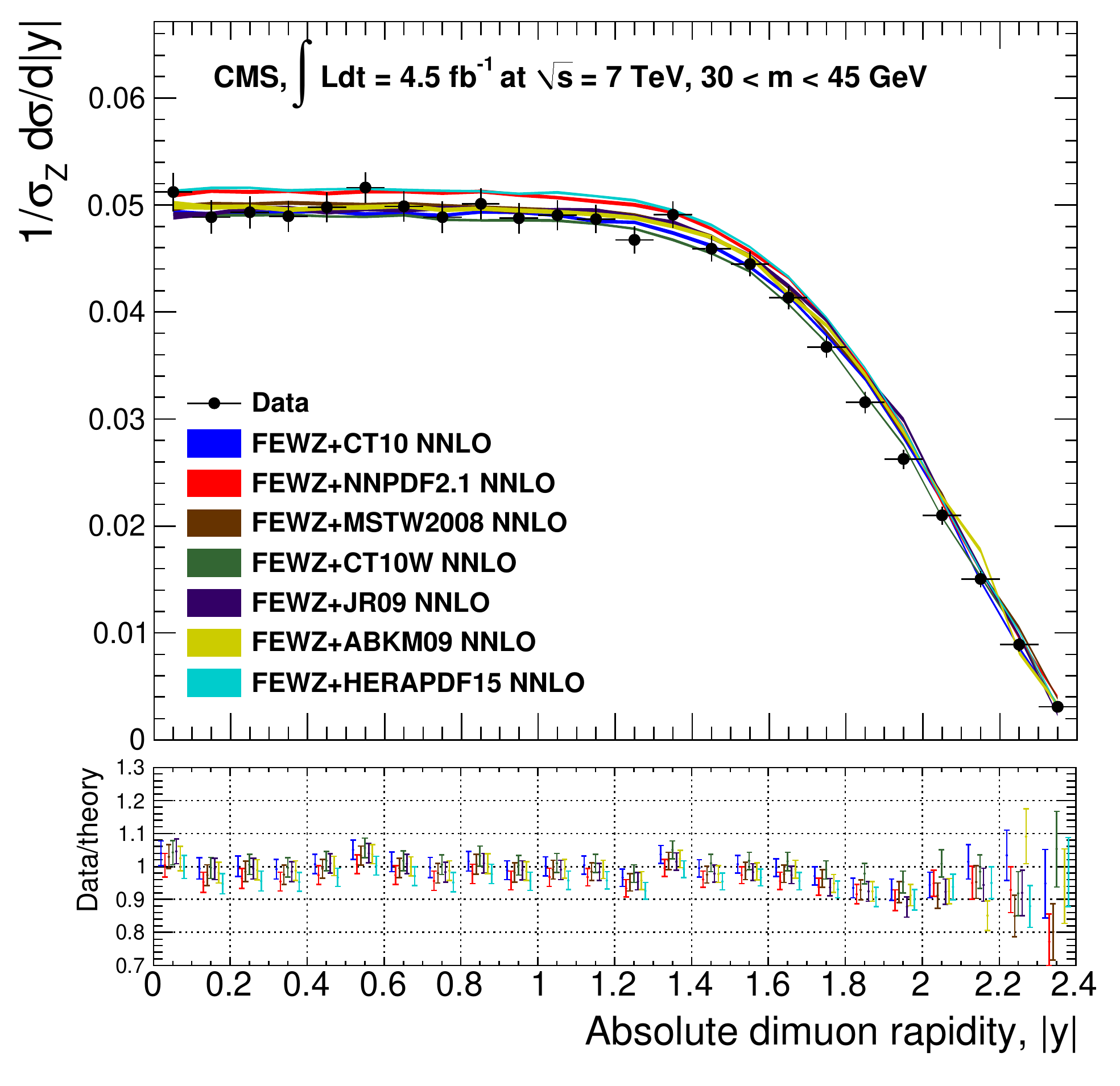}
   \end{center}
  \end{minipage}
 \end{tabular}
\caption{Double Drell-Yan differential cross section
as a function of dilepton mass and rapidity in two dilepton mass windows. Taken from Ref.~\cite{CMS03}.
 \label{fig:fig02}}
\end{center}
\end{figure}

  The CMS Collaboration measured the Drell-Yan distribution in electron,
muon and combined  channels
using 7 TeV data collected in the year 2011 ~\cite{CMS03}. The range of dilepton
mass is between 15 and 2000 GeV. Double differential cross section
as a function of dilepton mass and rapidity was measured 
only in muon channel in the restricted
mass range between 20 and 1500 GeV. The measured differential cross section normalised to the $Z$ cross section
is compared with the NNLO pQCD prediction combined with LO and NLO EW prediction calculated using FEWZ.
Both levels of EW predictions agree with the data.
The double differential cross section is compared with NNLO pQCD prediction using
various sets of PDFs (Figure~\ref{fig:fig02}). This is the first double differential Drell-Yan
cross section measurement performed at a hadron collider and will provide precise inputs to update PDFs.

  The CMS Collaboration also measured Drell-Yan differential and double differential cross section
using 19.7 fb$^{-1}$ of 8 TeV data ~\cite{CMS04}.
The range of dilepton mass is the same as in the 7 TeV measurement.
Results agree with the NNLO theoretical prediction calculated by the FEWZ program. 

  Both collaborations measured differential cross section of $Z$ boson as
a function of $Z$ transverse momentum and rapidity. The ATLAS Collaboration also investigated
the observable $\Phi^{*}_{\eta}$ - an alternative probe of the $Z$ transverse momentum, less sensitive
to the lepton energy and momentum uncertainties 
\footnote{:$  \Phi^{*}_{\eta}=\tan(\frac{\pi - \Delta\Phi}{2})\sin(\theta^{*}_{\eta})$, where
$\cos(\theta^{*}_{\eta})=\tanh((\eta^{-}-\eta^{+})/2)$ and $\Delta\Phi$ is the azimuthal opening angle between the two leptons~\cite{ATLAS07}.}.

   The ATLAS Collaboration measured normalised differential cross sections
of $Z$ using 4.7 fb$^{-1}$ of 7 TeV data~\cite{ATLAS04}. Cross section
was measured as a function of $Z$ transverse momentum in three rapidity
bins: $|y_{Z}|<1$, $1<|y_{Z}|<2$ and $2<|y_{Z}|<2.4$. The range of $Z$ boson transverse momentum is $<0,800>$ GeV.

\begin{figure}[!ht]
\begin{center}
 \begin{tabular}{cc}
  \begin{minipage}{.5\hsize}
   \begin{center}
     \includegraphics[width=7cm]{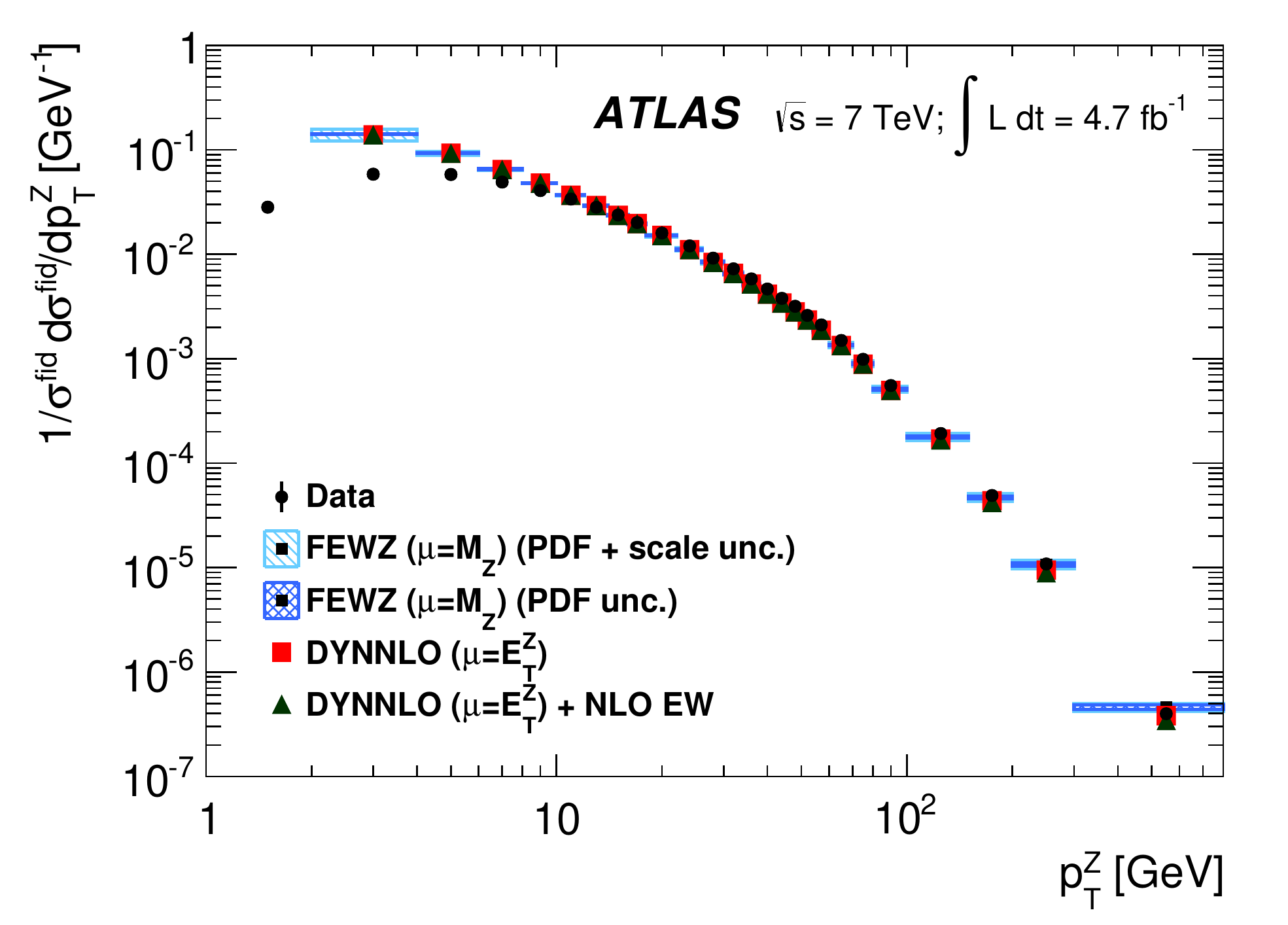}
   \end{center}
  \end{minipage}

  \begin{minipage}{.5\hsize}
   \begin{center}
     \includegraphics[width=7cm]{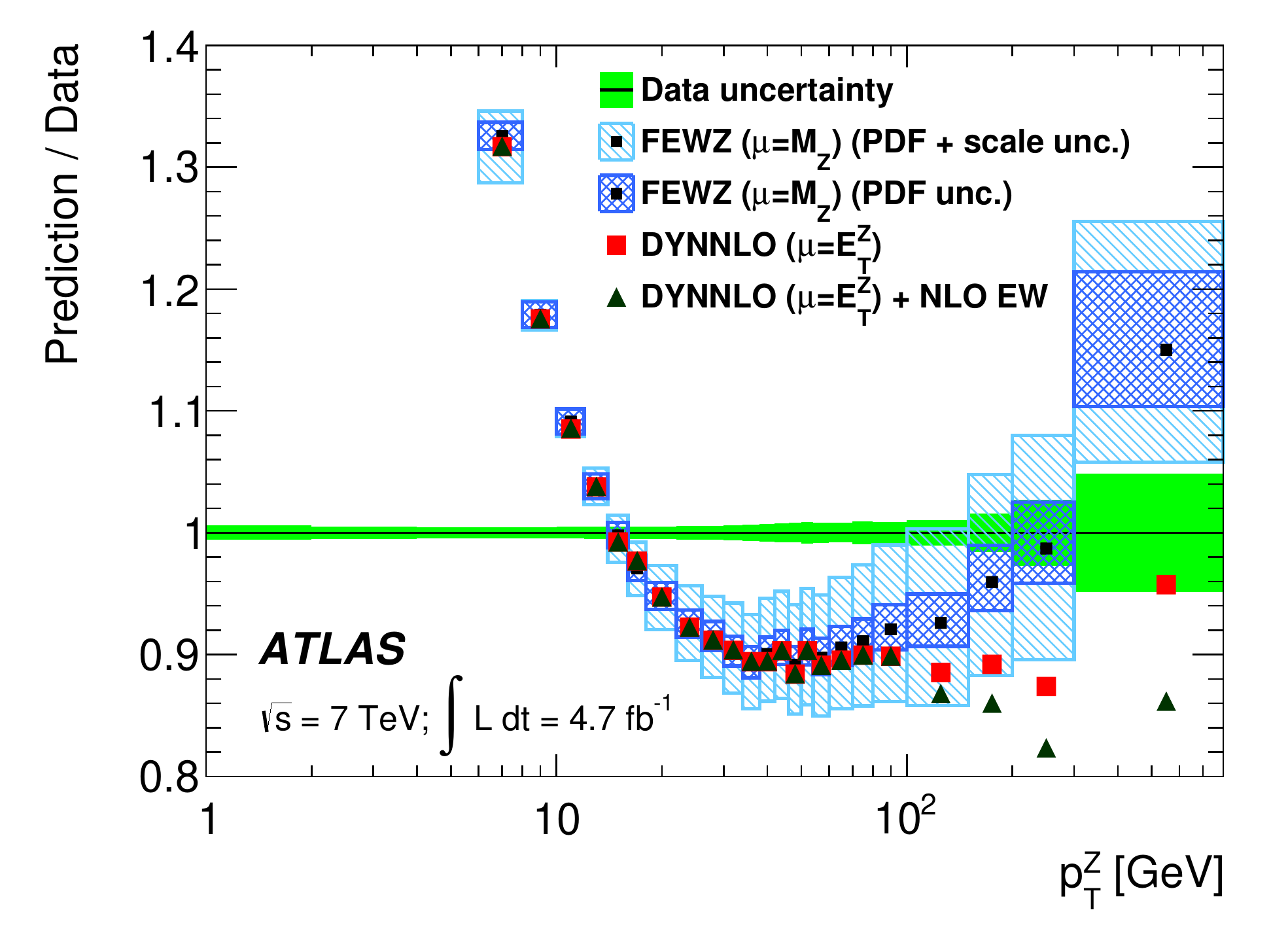}
   \end{center}
  \end{minipage}
\\
  \begin{minipage}{.5\hsize}
   \begin{center}
     \includegraphics[width=7cm]{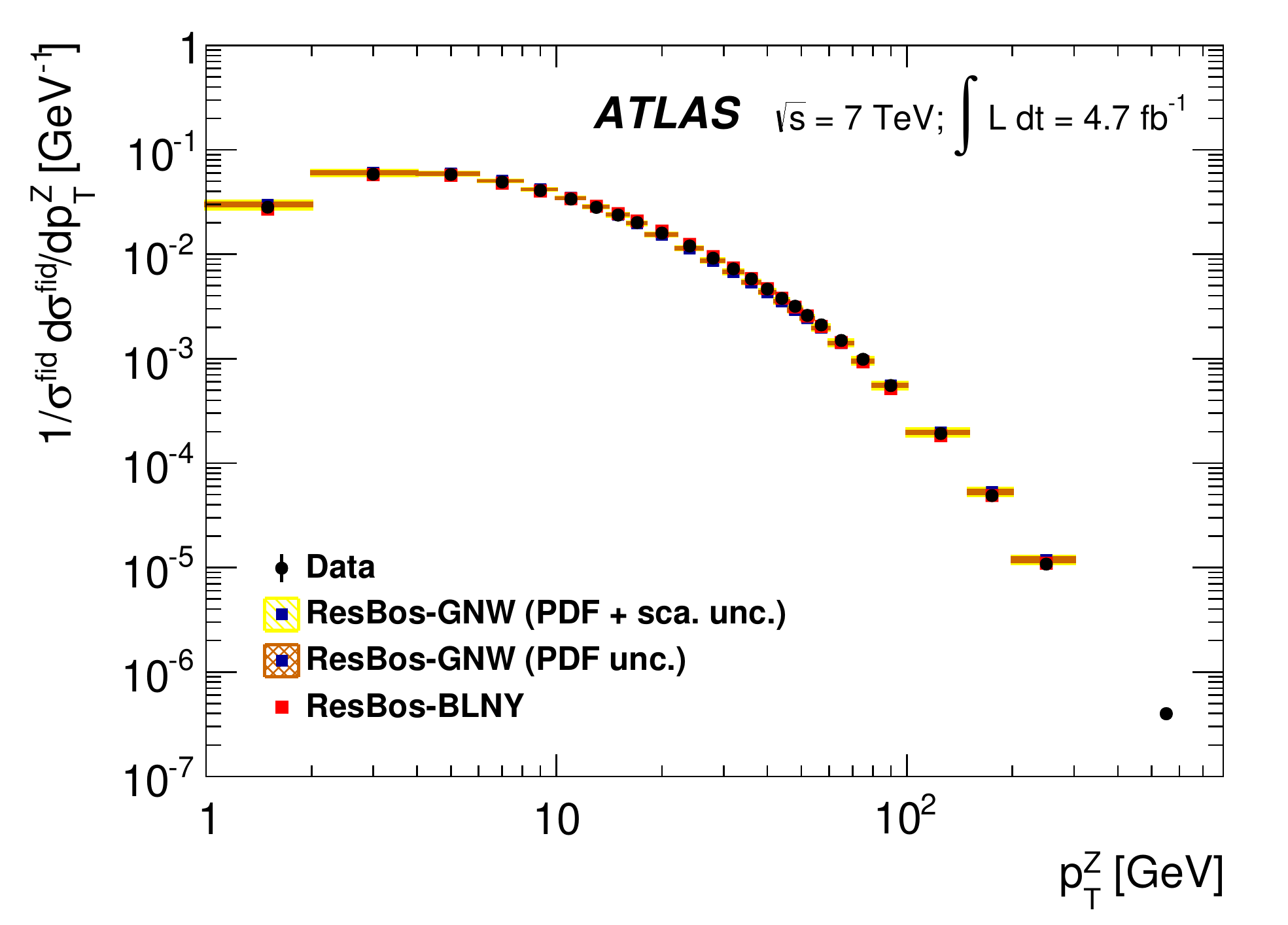}
   \end{center}
  \end{minipage}

  \begin{minipage}{.5\hsize}
   \begin{center}
     \includegraphics[width=7cm]{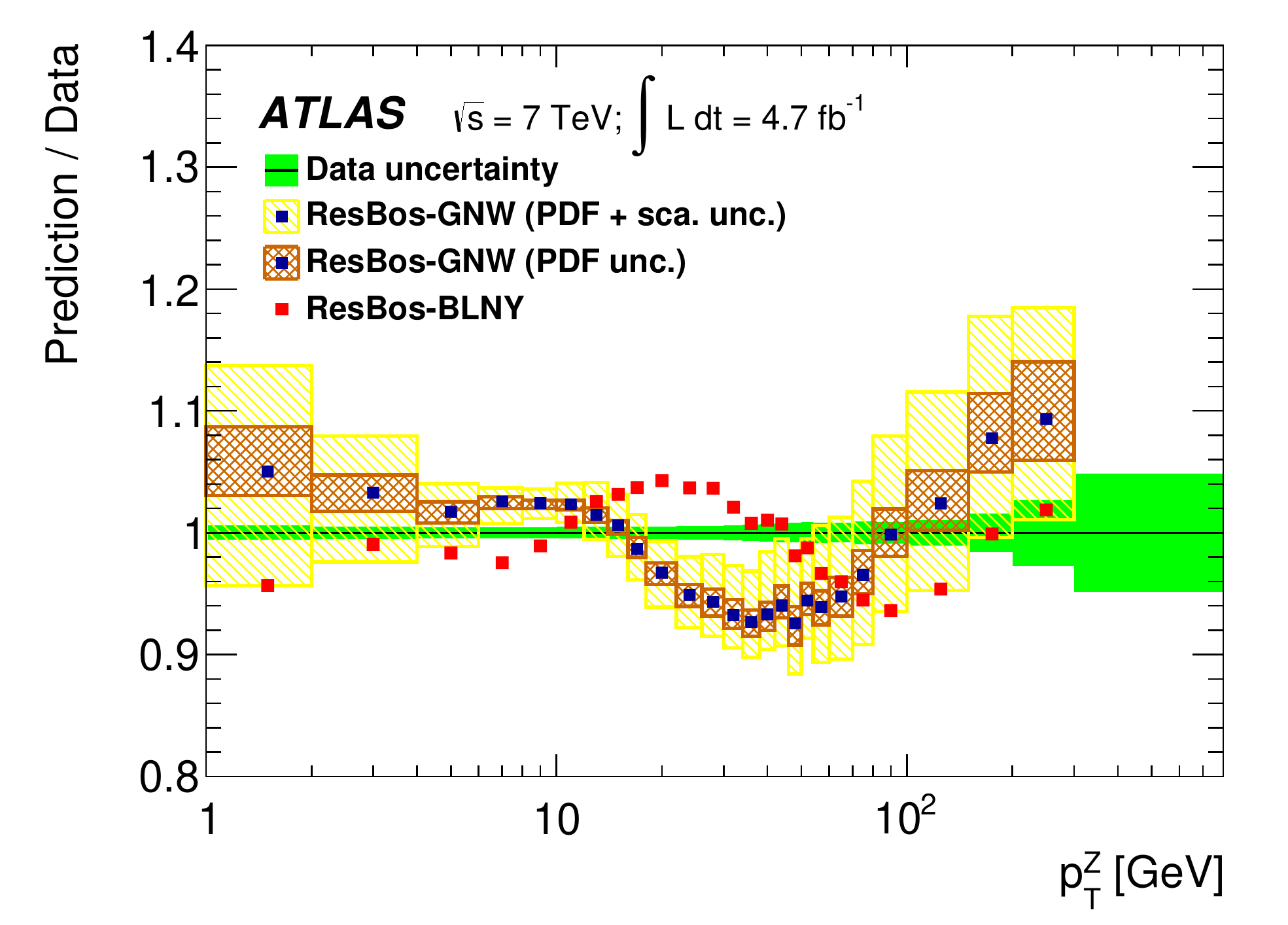}
   \end{center}
  \end{minipage}
 \end{tabular}
\caption{Top left: comparison of measured $P^{\scriptsize{\textrm{Z}}}_{{\scriptsize\textrm{T}}}$ distribution with fixed order NNLO calculations of FEWZ and DYNNLO.
Bottom left: comparison of measured $P^{\scriptsize{\textrm{Z}}}_{{\scriptsize\textrm{T}}}$ distribution to the soft gluon resummation calculation using RESBOS.
Right: ratios of theoretical predictions to data. Taken from Ref.~\cite{ATLAS04}.
 \label{fig:fig03}}
\end{center}
\end{figure}

  Comparison of theoretical predictions to the measured normalised 
differential cross-section is shown in Figure~\ref{fig:fig03}. As expected(two upper plots), fixed order NNLO predictions
using DYNNLO1.3~\cite{DYNNLO1},~\cite{DYNNLO2} and FEWZ can describe the data only at large transverse momenta
of $Z$ where radiation of high $p^{ }_{{\scriptsize\textrm{T}}}$ gluons dominates. Low $p^{ }_{{\scriptsize\textrm{T}}}$ region is dominated 
by soft gluon emission modeled by soft gluon resummation included 
in RESBOS~\cite{RESBOS1},~\cite{RESBOS2} (two lower plots).

  Measured normalised fiducial differential cross sections as a function
of transverse momentum and $\Phi^{*}_{\eta}$ observable were used for tuning event generators
PYTHIA8~\cite{PYTHIA8} in standalone mode and in a configuration interfaced to POWHEG.
  Tuned were the parameters affecting transverse momentum - ordered, interleaved parton shower
in PYTHIA8. 
The best prediction is provided by PYTHIA8, which is also able
to describe the different rapidity bins with the single tune.

  The CMS Collaboration measured absolute fiducial double differential cross section
of the $Z$ boson as a function of $Z$ transverse momentum and rapidity using 19.7
fb$^{-1}$ of 8 TeV data~\cite{CMS05}. The measurement
is done in dimuon channel only. The transverse momentum of $Z$ is in the range
$[0,300]$ GeV, the differential cross section is measured in six rapidity bins.

\begin{figure}[!ht]
\begin{center}
 \begin{tabular}{cc}
  \begin{minipage}{.3\hsize}
   \begin{center}
     \includegraphics[width=4cm]{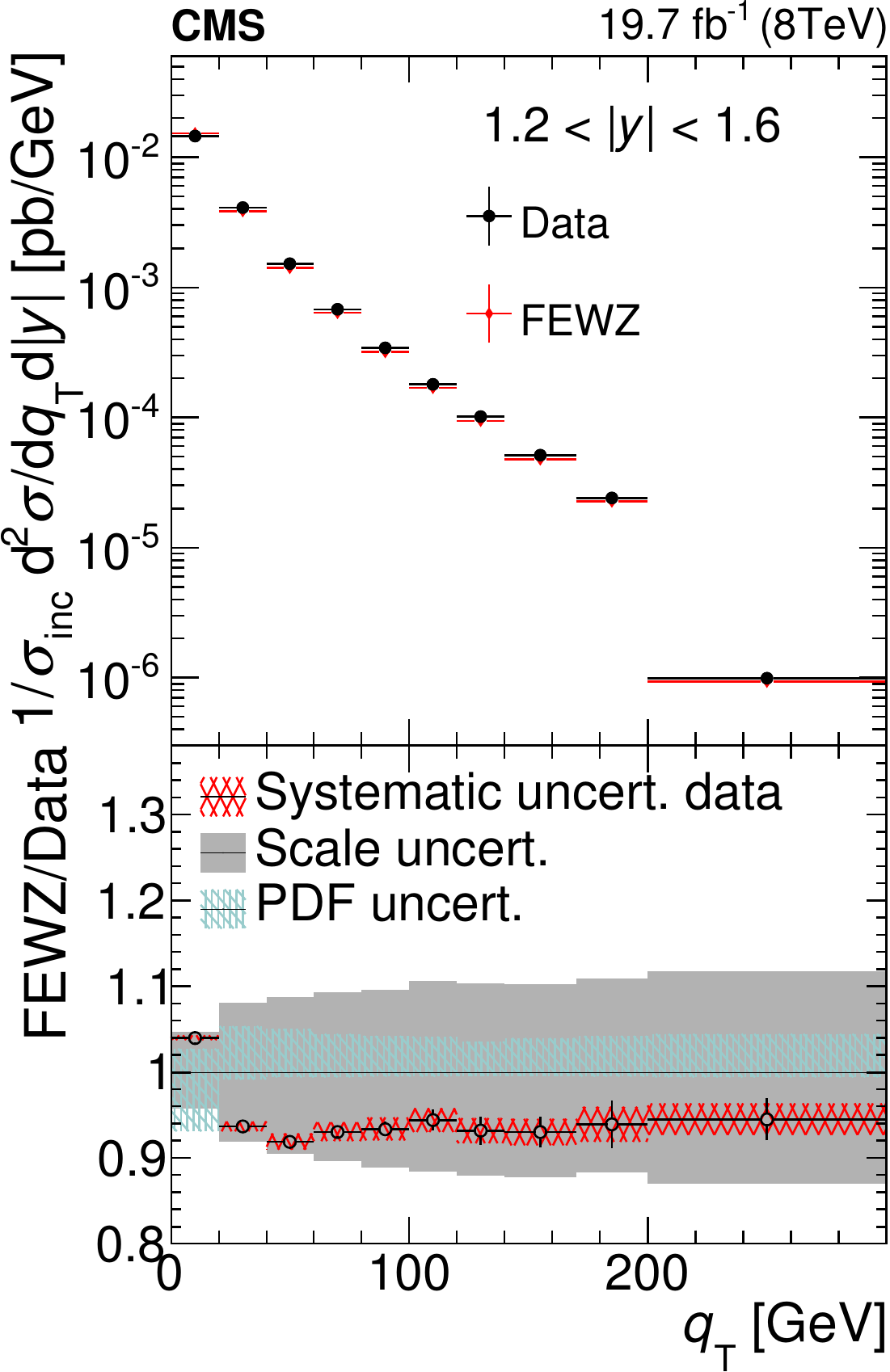}
   \end{center}
  \end{minipage}

  \begin{minipage}{.3\hsize}
   \begin{center}
     \includegraphics[width=4cm]{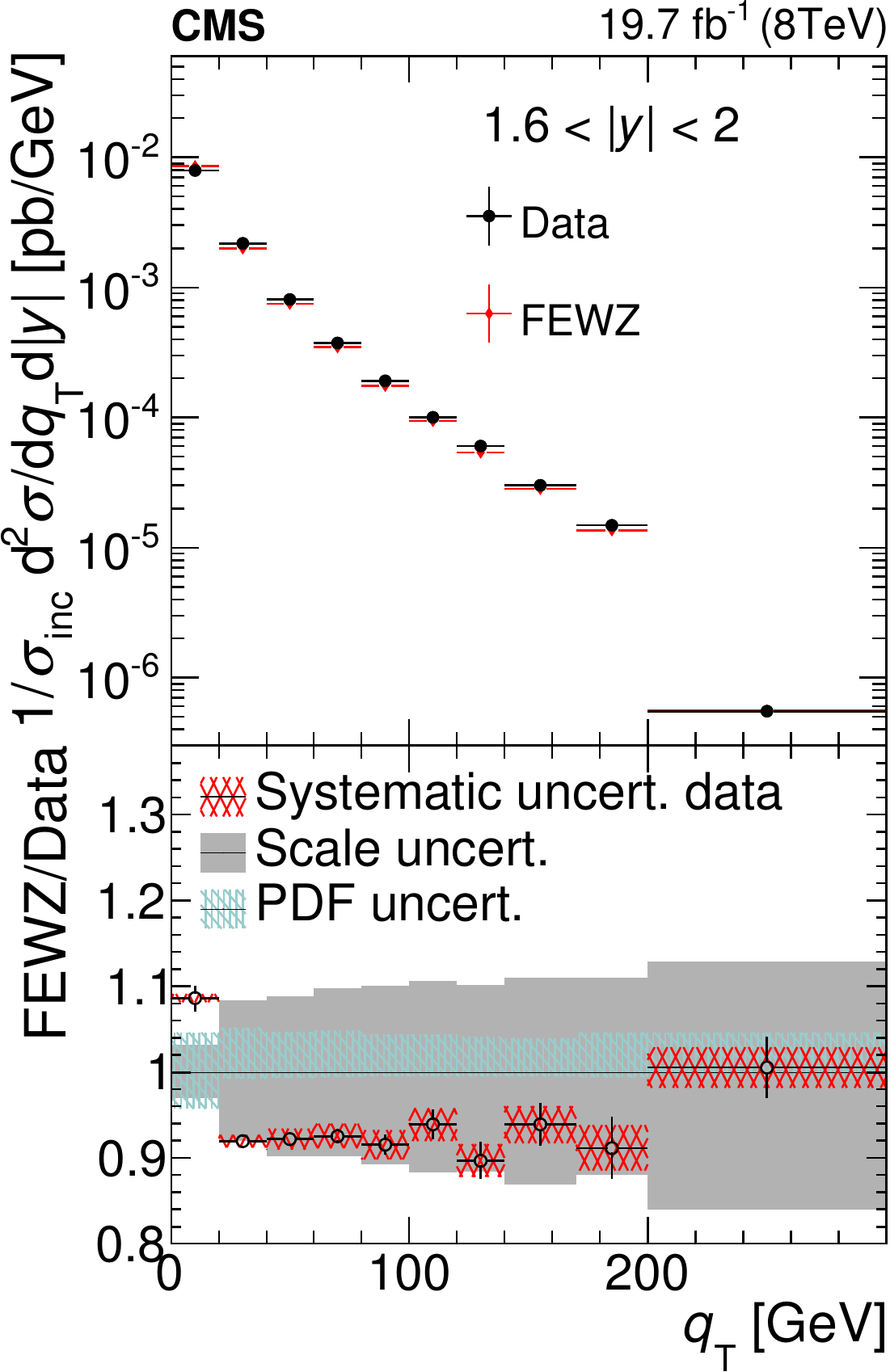}
   \end{center}
  \end{minipage}

  \begin{minipage}{.3\hsize}
   \begin{center}
     \includegraphics[width=4cm]{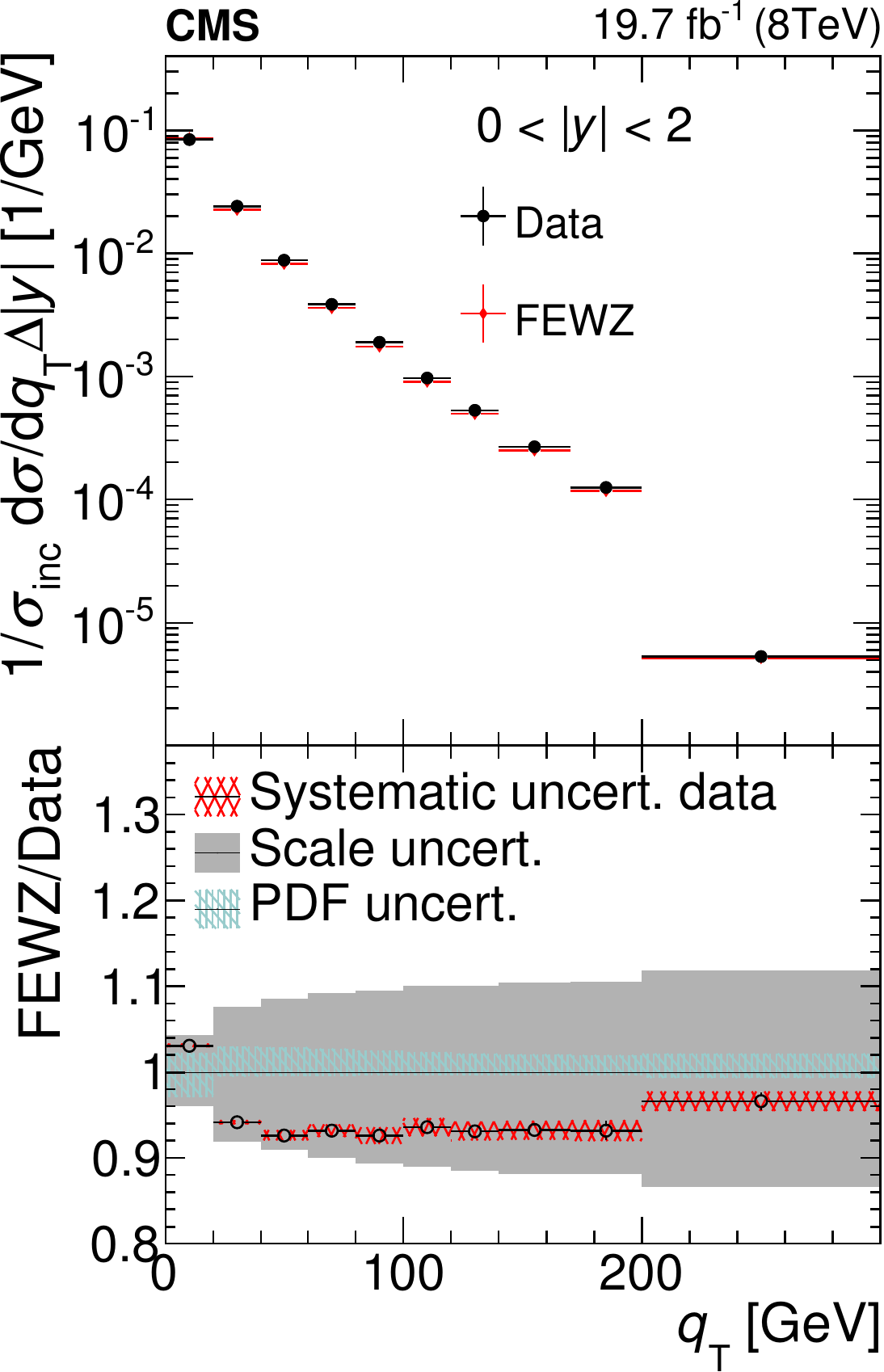}
   \end{center}
  \end{minipage}

 \end{tabular}
\caption{Comparison of measured $P^{\scriptsize{\textrm{Z}}}_{{\scriptsize\textrm{T}}}$ distribution with fixed order NNLO calculations performed by FEWZ.
Left two plots: $P^{\scriptsize{\textrm{Z}}}_{{\scriptsize\textrm{T}}}$ distributions in two selected rapidity bins. 
Right plot: $P^{\scriptsize{\textrm{Z}}}_{{\scriptsize\textrm{T}}}$ 
distribution integrated over the whole rapidity range. Taken from Ref.~\cite{CMS05}.
 \label{fig:fig04}}
\end{center}
\end{figure}

  The measured absolute fiducial $Z$ boson differential cross section
as a function of $Z$ transverse momentum is compared with the NNLO prediction
of FEWZ(Figure~\ref{fig:fig04}) in two selected rapidity bins (left two plots) and integrated over all rapidity bins
(right plot). Predictions agree with data within uncertainties, which are dominated 
by scale uncertainties. Starting from 20 GeV, the shape is predicted correctly.
For the lowest transverse momentum bin, the ratio prediciton/data is different from the
remaining bins - as expected for a fixed order prediction.

 Production properties of $W$ and $Z$ boson created in association with jets were measured
at 7 TeV and 8 TeV by the CMS Collaboration. The ATLAS Collaboration
published only measurements at 7 TeV.

  Using 19.6 fb$^{-1}$ of 8 TeV data~\cite{CMS06}, the CMS Collaboration
performed measurement of differential cross-section of $Z$ boson in association with
jets as a function of the jet multiplicity, of the transverse momentum and of absolute pseudorapidity
of the $n^{th}$ jet and of the scalar sum of the jet transverse momenta. Results of electron
and muon channels were combined.
Differential cross sections of the $1^{st}$ and $2^{nd}$ leading jets are compared to MADGRAPH~\cite{MADGRAPH} interfaced
with PYHTIA6~\cite{PYTHIA6} (LO) and to SHERPA~\cite{SHERPA} (NLO).
For the leading jet, MADGRAPH describes data very well instead of the region 150-450 GeV where an 
excess of events in data over prediction is observed.
Similar excess was observed by the CMS Collaboration at 7 TeV. SHERPA predicts slightly harder distributions
than the measurement. 

The ratio of the production cross section for $W$ and $Z$ bosons
in association with jets has been measured by the ATLAS Collaboration
using 4.6 fb$^{-1}$ of 7 TeV data~\cite{ATLAS05}.

Measurements of such ratios provide more precise test of pQCD
with respect the $W/Z$+jets measurements because of the cancellation of
some experimental uncertainties like luminosity and jet energy scale
and effects from non-perturbative processes (hadronization and
multiparton interactions).

\begin{figure}[!ht]
\begin{center}
 \begin{tabular}{cc}
  \begin{minipage}{.5\hsize}
   \begin{center}
     \includegraphics[width=7cm]{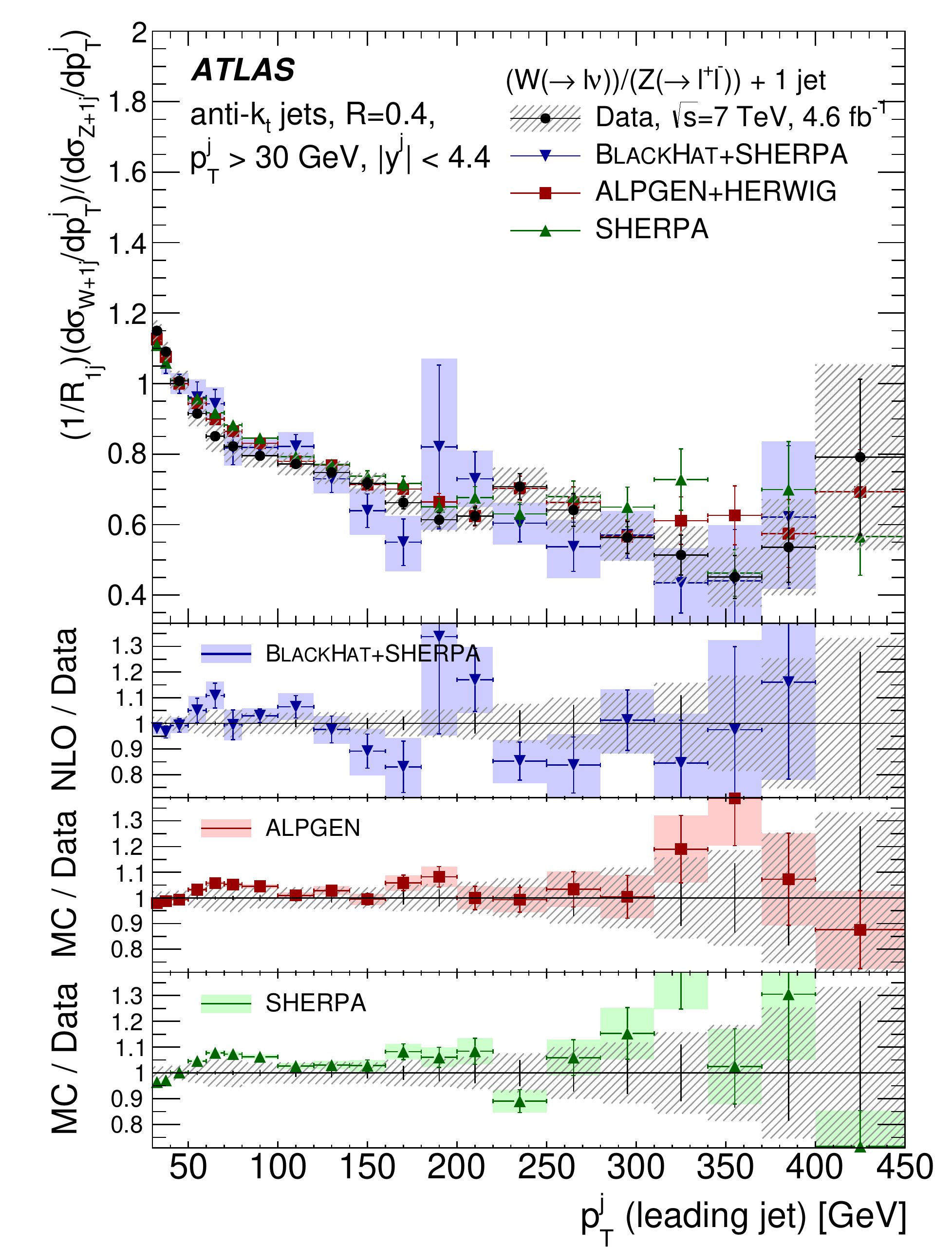}
   \end{center}
  \end{minipage}

  \begin{minipage}{.5\hsize}
   \begin{center}
     \includegraphics[width=7cm]{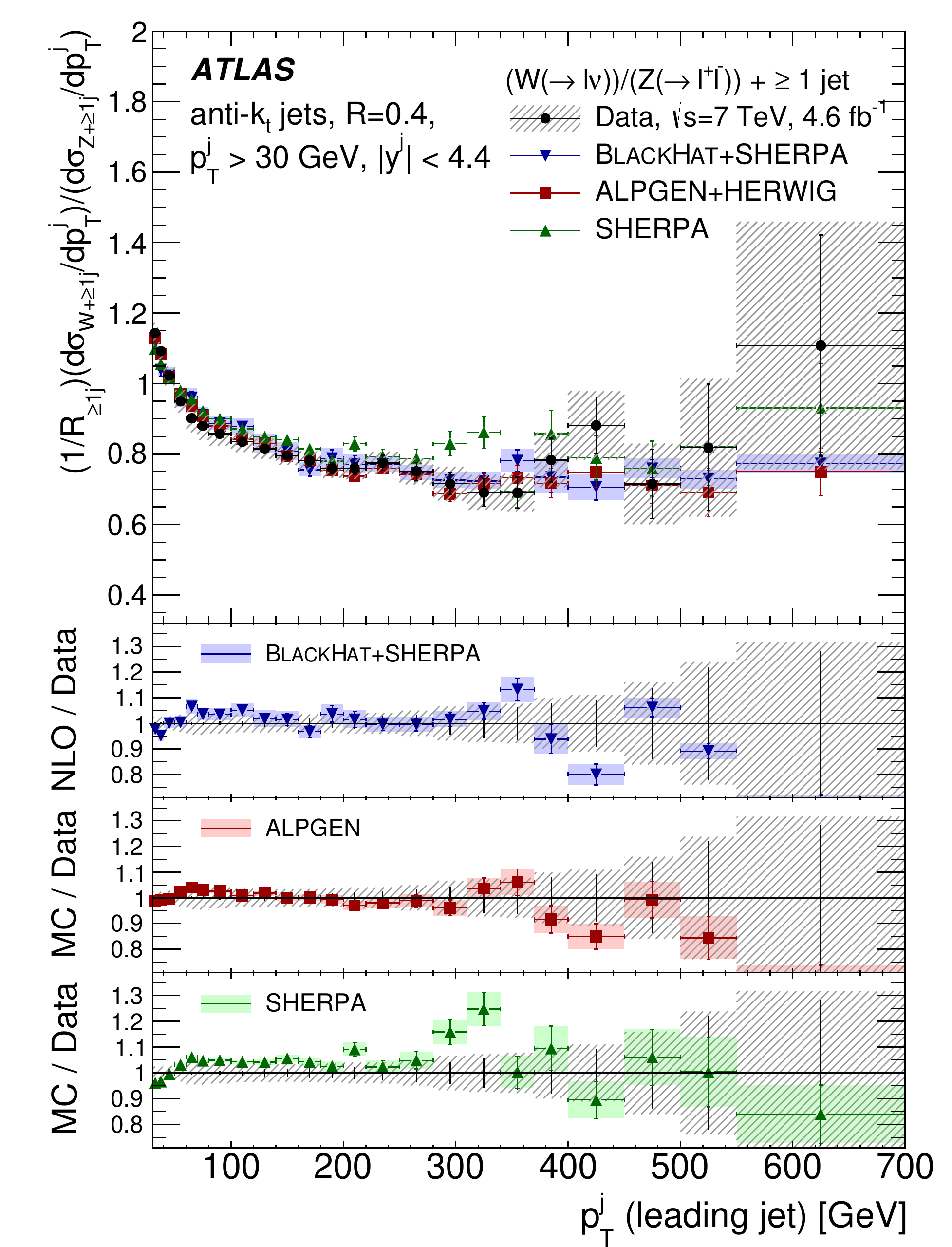}
   \end{center}
  \end{minipage}
 \end{tabular}
\caption{Ratio of $W$ + jets to $Z$ + jets production cross sections as a function
of the transverse momentum of the leading jet. Left: events with just 1 jet. Right: events with number of
jets $\geq$ 1. Taken from Ref.~\cite{ATLAS05}
 \label{fig:fig05}}
\end{center}
\end{figure}

Measurements are compared to NLO pQCD calculations using BLACKHAT + SHERPA~\cite{BLACKHAT}
and to preditions of ALPGEN~\cite{ALPGEN} and SHERPA LO event generators with parton showers.
All signal samples created by event generators were normalised to 
NNLO in pQCD using FEWZ.
Ratio of $W$ + jets to $Z$ + jets  cross sections was investigated as a function of the
exclusive jet multiplicity. Theoretical predictions describe data well,
with a few exceptions. SHERPA overestimates data in the region of 
high jet multiplicity.
Ratio of $W$ + jets to $Z$ + jets production cross sections as a function
of the transverse momentum of the leading jet for events with just 1 jet(left plot)
and for events with 1 and more jets(right plot) is 
shown in Figure~\ref{fig:fig05}.
  The drop in the ratio in the low $p^{ }_{{\scriptsize\textrm{T}}}$ region is due to the $W$ and $Z$ bosons mass difference,
which affects the scale of parton radiation, as well as due to the different vector 
boson polarization affecting kinematics of their decay products.
ALPGEN provides the best description of data.

  Measurements of production of massive vector bosons in association 
with the $c$ and $b$ quarks were performed by both collaborations.

The ATLAS Collaboration measured differential production cross sections
for a $Z$ boson decaying into an electron or muon pair
in association with $b$ jets using 4.6 fb$^{-1}$ of 7 TeV data~\cite{ATLAS06}.

\begin{figure}[!ht]
\begin{center}
 \begin{tabular}{cc}
  \begin{minipage}{.5\hsize}
   \begin{center}
     \includegraphics[width=7cm]{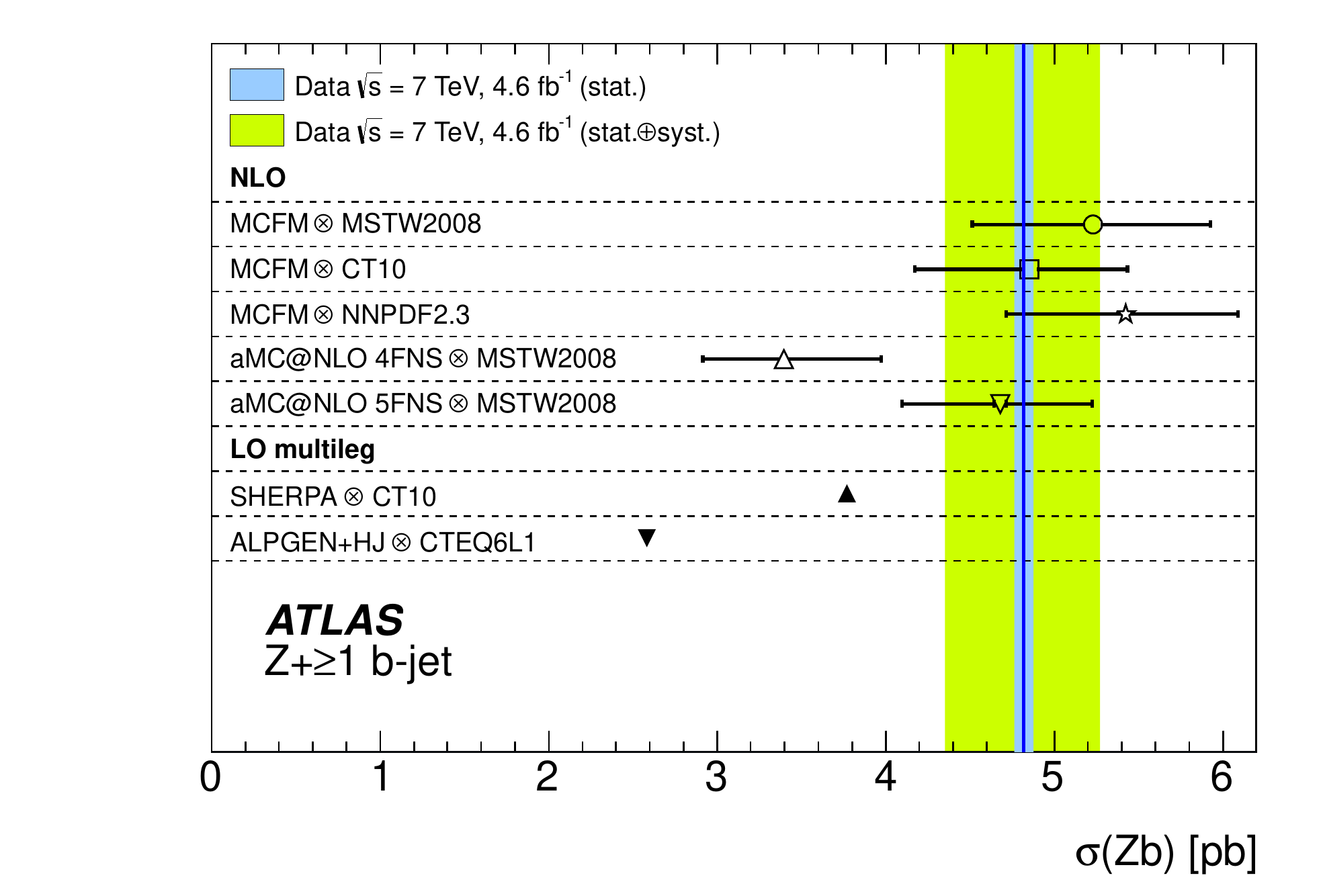}
   \end{center}
  \end{minipage}

  \begin{minipage}{.5\hsize}
   \begin{center}
     \includegraphics[width=7cm]{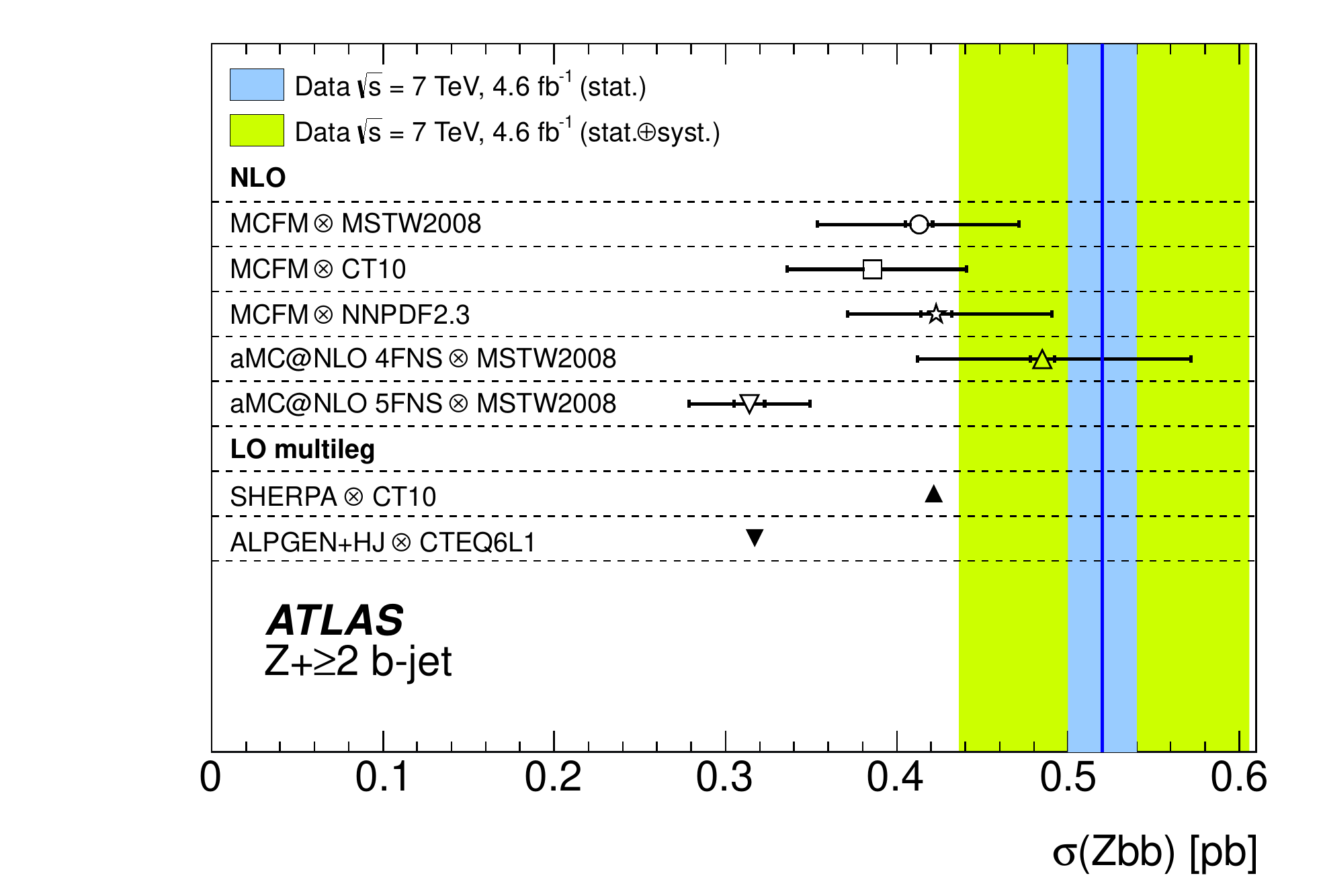}
   \end{center}
  \end{minipage}
 \end{tabular}
\caption{Left: cross sections for $Z$ + $\geq$ 1 $b$-jet. Right: cross sections for $Z$ + $\geq$ 2 $b$-jets. Taken from Ref.~\cite{ATLAS06}.
 \label{fig:fig06}}
\end{center}
\end{figure}

Cross sections for $Z$ + $\geq$ 1 $b$-jet (left plot) and for $Z$ + $\geq$ 2 $b$-jets (right plot)
are shown in Figure~\ref{fig:fig06}. The best description of data is provided by
the NLO prediction of parton-level integrator MCFM~\cite{MCFM}. The underestimation
of the $Z$ + $\geq$ 2 $b$-jet cross section by the aMCatNLO~\cite{AMCATNLO} using the five flavour
number scheme is probably due to the fact that it uses only LO prediction
for $\geq$ 2 $b$-jet $b$-jet process.

The differential cross section as a function of a $b$-jet transverse momentum and
rapidity is best described by the aMCatNLO prediction using the five flavour number
scheme. The same is true for the case of the differential cross section as a function
of $Z$ transverse momentum and rapidity.

\newpage

\section*{Summary}

During the Run 1 period, the ATLAS and CMS Collaborations produced nearly 60 publications dealing with
$W/Z$ physics in total. Typical precision of published values of integrated $W$ and $Z$ cross
sections is at the level of a few percent, both for 7 TeV and 8 TeV data. 
Predictions of Standard Model were tested in most cases at NLO accuracy or higher.
Analysis of data taken in the years 2010 and 2011 is nearly completed.
Analysis of 8 TeV data from the year 2012 is still ongoing.

  During the Run 2 period, center of mass energy will increase by a factor of 1.6. A new kinematic regime
will be encountered. According the current expectations, about 100 fb$^{-1}$ of data will be collected during
the whole Run 2 period, what represents four times the amount collected during Run 1. 
About 10 fb$^{-1}$ is expected during the first year of data taking. This constitutes
14 millions of $W^{-}$, 20 millions of $W^{+}$ and 4 millions of $Z$.
These data will be used for basic tests of the Standard Model in the new kinematic region.

\end{document}